\documentclass[nonacm, ]{acmart}

\usepackage{xcolor}
\definecolor{DarkGreen}{HTML}{000000}

\usepackage{graphicx}
\usepackage{caption}
\usepackage{subcaption}
\usepackage{todonotes}
\usepackage{multirow}
\usepackage{colortbl}
\definecolor{lightgray}{gray}{0.9}
\AtBeginDocument{%
  \providecommand\BibTeX{{%
    \normalfont B\kern-0.5em{\scshape i\kern-0.25em b}\kern-0.8em\TeX}}}
\begin{document}

\title[College Students' Experiences Navigating the Higher Education Environment in a Generative AI World]{``Everyone's using it, but no one is allowed to talk about it'': College Students' Experiences Navigating the Higher Education Environment in a Generative AI World}

\author{Yue Fu}
\email{chrisfu@uw.edu}
\affiliation{%
  \institution{Information School, University of Washington}
  \city{Seattle}
  \state{Washington}
  \country{US}
}

\author{Yifan Lin}
\affiliation{%
  \institution{Information School, University of Washington}
  \city{Seattle}
  \state{Washington}
  \country{US}}
\email{}

\author{Jessica Wang}
\affiliation{%
  \institution{Information School, University of Washington}
  \city{Seattle}
  \state{Washington}
  \country{US}}
\email{}

\author{Sarah Tran}
\affiliation{%
  \institution{Information School, University of Washington}
  \city{Seattle}
  \state{Washington}
  \country{US}}
\email{}

\author{Alexis Hiniker}
\affiliation{%
  \institution{Information School, University of Washington}
  \city{Seattle}
  \state{Washington}
  \country{US}}
\email{}

\begin{abstract}
Higher education students are increasingly using generative AI in their academic work. However, existing institutional practices have not yet adapted to this shift. Through semi-structured interviews with 23 college students, our study examines the environmental and social factors that influence students' use of AI. Findings show that institutional pressure factors like deadlines, exam cycles, and grading lead students to engage with AI even when they think it undermines their learning. Social influences, particularly peer micro-communities, establish de-facto AI norms regardless of official AI policies. Campus-wide ``AI shame'' is prevalent, often pushing AI use underground. Current institutional AI policies are perceived as generic, inconsistent, and confusing, resulting in routine noncompliance. Additionally, students develop value-based self-regulation strategies, but environmental pressures create a gap between students' intentions and their behaviors. Our findings show student AI use to be a situated practice, and we discuss implications for institutions, instructors, and system tool designers to effectively support student learning with AI.
\end{abstract}



\maketitle

\section{Introduction}

AI is rapidly transforming higher education. Since the public release of ChatGPT in late 2022, multiple tools such as Perplexity, Gemini, Claude and the like have shaped how students learn \cite{schei2024perceptions}, collaborate with peers \cite{exploring, perifanou2025collaborative}, and interact with instructors \cite{billy2023study}, as well as how faculty design teaching and assessment \cite{lau2023ban, lo2023impact, rudolph2023chatgpt}. Students across higher education institutions worldwide have rapidly adopted AI tools, with multiple recent surveys showing 60-86\% of students using AI for academic work \cite{fovsner2024university, hu2025status, rong2024digital}, a dramatic increase from 43\% just two years prior \cite{welding_2023}.

Existing work describes some of the reasons \textit{why} students turn to AI (e.g., to clarify complex concepts, manage workload, and improve performance \cite{morell2025characteristics}), which assumes students make autonomous decisions about AI based on individual preferences and motivations. However, we know far less about the \textit{environment} in which college students are making those choices. By environment we mean those factors outside a learner’s own cognition, affect, or skills such as assignment deadlines and exam schedules, grading regimes, institutional rules, social norms, and AI policies on campus. Pre-AI research on academic misconduct already shows that such environmental factors (heavy workload, assessment timing, instructor practices, peer influence) shape student academic 
behavior \cite{miles2022students, wiley_2022}. Without understanding these environmental factors, higher education institutions cannot develop effective responses to AI integration in higher education and design effective AI policy. 

We examine how environmental factors on campus drive AI adoption and engagement, and how students adapt through self-regulation. We focus on how students perceive and navigate these factors that arise from the study environment with generative AI, and we analyze how students set personal boundaries based on their academic values in response. Specifically, we asked these three research questions: 
\vspace{2mm}

\textbf{RQ1:} What environmental factors influence students' engagement with AI in academic contexts?

\textbf{RQ2:} How do students feel about these influences?

\textbf{RQ3:} What AI-related institutional and policy changes do students recommend?

\vspace{2mm}

To answer these questions, we interviewed 23 students in undergraduate, master's, and doctoral programs at a large public university in the United States. These semi-structured interviews asked students how they navigated higher education with generative AI and probed how students respond to these environmental factors. In addition, we solicited their recommendations for institutional changes to better facilitate and support their learning in a generative AI world.

Our findings show that institutional and social factors significantly shaped students’ AI usage, perception, and engagement. Specifically, deadlines, examination cycles, and grading schemes influenced both the frequency and pattern of AI use, leading students to engage with AI even when they think it undermines their learning. Additionally, students' immediate social groups established de facto norms about AI usage, often overriding official policy guidelines. We discovered there is widespread ``\textit{AI shame}'' phenomenon on campus, characterized by a culture where ``\textit{everybody uses it, but nobody talks about it}.'' Students also perceived existing AI policies on campus as generic, outdated, and inconsistent, and they reported that it is common for students to violate them.

Most students described being aware of the potential negative effects of relying on AI, prompting them to adopt various personal boundaries and self-regulation strategies aligned with their values. Despite these intentions, students often reported a noticeable gap between their ideal standards for themselves and actual behaviors (an ``\textit{intention-behavior gap}''), particularly when faced with environmental pressures.

Students argued that blanket bans on AI are impractical and likely to push usage underground. Instead, they recommended creating AI policies collaboratively between students and instructors to promote greater transparency, understanding, and compliance. They also proposed increasing in-person evaluations such as quizzes and exams to effectively assess learning in an environment influenced by AI. Finally, students showed a lack of awareness about AI resources on campus, advocating strongly for AI-literacy classes and clear guidelines to help students integrate AI responsibly into their study.

This research contributes to the community by:

\begin{itemize}

    \item Identifying environmental factors---including institutional pressures and social influences---that shape students' AI usage. Based on these, we propose institutional changes to support student learning with AI in higher education.
    \item Revealing student perceptions of current AI policies and common noncompliance. We offer policy recommendations based on students' suggestions that align with their actual practices.
    \item Examining students' self-regulation strategies in response to environmental pressures. We suggest ways in which educators and AI-tool designers can support the responsible use of AI by students.
    \item Discussing how higher education institutions, instructors, and system designers could do to support students' learning in this generative AI world.

\end{itemize}

\section{Related Work}

\subsection{Students’ Usage, Perceptions, and Motivations toward Generative AI in Higher Education}
Students in higher education have rapidly adopted generative AI tools for academic work, reflecting a global trend \cite{sublime2024chatgpt, fovsner2024university, von2023artificial}. Surveys consistently report high usage rates, with students worldwide already using AI regularly in their academic activities \cite{fovsner2024university, von2023artificial, salih2024perceptions, ravvselj2025higher, hu2025status, rong2024digital, chan2023students}. For example, a 2024 survey aiming to understand AI usage of Slovenia university students across various academic levels and disciplines show 73\% of students use AI tools often or very often and 89\% students have a positive attitude towards AI tools in education. Only two percent of students do not use AI tools. And 93\% agreed that AI tool usage is common among peers \cite{fovsner2024university}. Students use AI for a variety academic tasks including supporting concept understanding, grammar and editing, information search and reading, brainstorming ideas \cite{von2023artificial, sarwanti2024they, rong2024digital}. Usage patterns vary significantly by demographic group, with male, STEM, and non-native-speaking students typically demonstrating higher confidence and greater engagement with AI tools \cite{freeman2025student, baek2024chatgpt}. 

Students largely have positive perception towards using AI \cite{rodway2023impact, chan2024students, kim2025examining, ravvselj2025higher, salih2024perceptions}. Studies mention benefits such as increased efficiency and productivity \cite{fovsner2024university, chauke2024postgraduate, sarwanti2024they, salih2024perceptions}, personalized support \cite{zhu2023harness, zhou2024unveiling, chan2023students, sarwanti2024they}, accessibility \cite{kuykendall_2023}, supporting idea generation \cite{zhu2023harness, sarwanti2024they}, increasing confidence \cite{ravvselj2025higher}, and improving language capability for non-native speakers\cite{zhu2023harness}. At the same time, students are cautious about AI usage and voice concerns and skepticism. Key concerns include the risk of academic misconduct and plagiarism \cite{fovsner2024university, salih2024perceptions} , inaccuracies and AI hallucination \cite{freeman2025student}, and the potential for overreliance on AI tools \cite{mogavi2024chatgpt, zhou2024unveiling}. Students also voice broader ethical worries such as data privacy \cite{zhou2024unveiling, balogh2024educational}, originality and authorship \cite{rudolph2023chatgpt}, cultural sensitivity of AI content \cite{salih2024perceptions}. 

Faculty and students both worry that reliance on AI could diminish foundational learning competencies, including critical thinking \cite{sarwanti2024they, mogavi2024chatgpt, kim2025examining}. Many educators fear that if students use AI to do the heavy cognitive lifting (writing essays, solving problems), they might learn less deeply. A US faculty and student survey found  both groups tended to agree generative AI would have ``more negative effects on foundational learning competencies, such as problem-solving, critical thinking, teamwork, self-efficacy, and motivation \cite{kim2025examining}. However, compare to faculties, students were significantly more positive than about AI's potential benefits \cite{kim2025examining}.

Studies exploring student motivation identify primary drivers such as pursuit of better understanding and academic performance, efficiency, and curiosity of new learning approaches \cite{morell2025characteristics, hasanein2023drivers}. Another study shows novelty and entertainment were strong motivations for college students' early adoption, while sustained academic use correlates strongly with practical needs such as seeking study guidance \cite{lee2025motivations}. One survey study analyzes social influence showing peer popularity is a powerful driver of ChatGPT adoption among undergraduates, while graduates rely more on autonomy and task relevance \cite{korchak2025role}. Despite insights into student motivations, there remains limited qualitative research examining social and institutional factors shape student AI adoption and usage. 

This paper builds on existing literature by situating student AI engagement within the broader institutional and social context, examining how these environmental factors shape their academic use of generative AI.

\subsection{Institutional AI Policies and Student Perspectives}

As generative AI use rapidly expands on college campuses, research increasingly reports the need for effective institutional policies and guidelines \cite{sarwanti2024they, liu2024qualitative}. There is a broad consensus that unregulated AI use poses risks to the fairness and academic integrity, and both students and faculty are calling for ethical guidelines to address this issue \cite{ravvselj2025higher, chauke2024postgraduate, zhou2024unveiling}. Despite this consensus, higher education institutions remain at an early stage of developing comprehensive, practical policies to manage AI integration \cite{zhou2024unveiling}. Both students and faculty recommend clear institutional guidelines and structured AI literacy training \cite{kim2025examining, ravvselj2025higher}.

Existing AI policies from leading universities predominantly emphasize concerns such as the originality of student work, reliability of AI-generated outputs, and equitable access to AI resources \cite{luo2024critical, Sinha_Burd_Preez_2023}. Analyses reveal that current policies often narrowly frame AI use within traditional academic misconduct frameworks, and bans on generative AI are unsustainable as student adoption increases \cite{an2025investigating}. Furthermore, research highlights that most institutional guidance targets faculty rather than students or broader stakeholders \cite{an2025investigating}. Notably, student perspectives frequently remain missing from institutional conversations around AI policy development \cite{crosscampus, yang2025undergraduate}. Research thus calls for clearer definitions of originality in student work, increased ethical consideration, transparency, interdisciplinary collaboration, and evidence-based policy development \cite{luo2024critical, bond2024meta, adnin2025examining}. 

Collectively, much of the existing research approach is based on quantitative surveys \cite{von2023artificial, barrett2023not, fovsner2024university, salih2024perceptions, ghimire2024generative, chan2023comprehensive, chan2024will} or policy content analyses \cite{humble2025higher, erhardt2025policy, luo2024critical}, scarce qualitative studies \cite{ simkute2025new} probe accounts from students explaining their underlying experiences, rationales, and perception towards institutional policies and environmental factors in general. Our study fills this gap by probing qualitative accounts from students about current situated in higher education environment. We report students’ perceptions and recommendations and discuss the implication for institutions, educators, and AI tool developers.

\section{Methods}
We conducted 23 semi-structured interviews with bachelor’s, master’s, and PhD students from a large public university in the United States.

\subsection{Participants}

We recruited participants between May and July 2025 through academic Slack and Discord channels and flyers posted in campus buildings across various departments. 
Additionally, the first author briefly mentioned the recruitment opportunity in a class the author taught and shared the information with other instructors, who informed their students the participation opportunity. Participants first completed an initial Qualtrics survey collecting demographic data (e.g., major, year of study), the specific AI tools used (e.g., ChatGPT, Perplexity, Claude), and descriptions of typical AI usage cases. Eligible participants were currently enrolled students who had some experience with at least one generative AI tool, a criterion met by all individuals who signed up.

Participants received a \$40 Amazon gift card upon completing the interview. Recruitment occurred in iterative waves, concurrent with ongoing data analysis. We analyzed data during biweekly research meetings and concluded that saturation was reached after interviewing approximately 18 participants  (see section  \ref{analysis} for detail). We stopped recruiting then, however, five additional participants had already been scheduled at that point, so we completed those interviews, bringing our total to 23 participants. Participant table is summarized in Appendix A, Table \ref{tab:participant_demographics}. 

\subsection{Materials and Procedure}

Three researchers first individually drafted a preliminary set of open-ended interview questions based on the research questions. The research team then collaboratively refined the interview protocol, focusing on understanding students’ motivations, perceived value of AI, the role of environmental factors (e.g., academic calendars, assignment deadlines), the influence of AI on learning, and suggestions for institution and policy changes. Then one researcher summarized the interview questions into the first draft of interview protocol. A pilot interview was conducted to test the protocol, leading to adjustments in question wording, sequencing, and the inclusion of additional follow-up questions.

The final semi-structured protocol had five main sections. The first section asked descriptive questions about participants' general AI usage, including when they started and their typical use cases. The second asked students’ motivations of using or avoiding AI. In the third section, we asked environmental influences on AI use, specifically focusing on course structure, institutional policies, peer influence. The fourth section asked the impact of AI on students' learning experiences, along with their desired support for integrating AI effectively. Finally, participants provided recommendations for institutional AI policies and changes that would best support their learning. The interviews were flexible following participants’ lead, allowing interviewers to follow up on relevant threads introduced by participants. Questions were added or reordered depending on participants’ prior reflections and experiences.

Interviews lasted approximately 60 minutes (ranging from 48 to 78 minutes) and were conducted via Zoom. Before recording, we changed participant names on Zoom with assigned participant IDs for anonymity. Interviews were audio-recorded and transcribed verbatim. Our institutional review board (IRB) reviewed and granted an exemption for the study.

\subsection{Data Analysis}
\label{analysis}
We employed Reflexive Thematic Analysis \cite{braun2019reflecting, braun2021one} to analyze interview transcripts. This method facilitates flexible, in-depth exploration of student experiences while explicitly acknowledging the researchers’ active and interpretative roles. Analysis began concurrently with data collection. Team members read transcripts and listened to audio recordings. Four researchers conducted preliminary analyses by each analyzing four different interview transcripts. During biweekly team meetings, we reviewed preliminary findings, shared insights, and discussed interpretations of key themes and representative quotes.

Through iterative discussions, we identified major themes related to research questions influencing students’ AI adoption, usage, and perceptions. In early July 2025, we conducted an internal workshop using the collaborative tool Figma \cite{figma_2025} to organize and cluster representative quotes into thematic categories. During this workshop, we agreed that we had reached data saturation, as consistent patterns and themes repeatedly emerged across participants. This consensus made us decide to conclude participant recruitment.

Subsequently, two researchers individually divided and re-examined all transcripts, extracting quotes, refining and integrating themes through frequent discussions. Next, one researcher organized participants' quotes and themes using a qualitative analysis research tool Delvetool \cite{delve_2019} and then compiled into a comprehensive document. The researcher further condensed, refined, and discard themes not directly relevant to the research questions. The researcher then conducted a second iteration of this organized document, identifying most representative quotes and updating names of each theme for writing purpose. Subsequently, the lead author drafted the initial version of the Results section based on this final thematic organization, which the research team collaboratively edited it. 

\section{Results}

\subsection{The Influence of Academic and Institutional Pressure}

Students said that campus institutional and environmental factors beyond their personal cognition and control influenced their engagement with AI. They constantly were ``\textit{running against time to use AI}'' (P22), and mentioned looming deadlines, exam cycles, grading policy, and the challenge of juggling multiple coursework as significant drivers of AI use. These institutional pressures shaped both students' frequency and usage patterns.

\subsubsection{The High Baseline Stress on Campus Makes AI Feel Necessary} \label{stress}
Across participants, they mentioned baseline stress in college as an ubiquitous background that shaped their AI usage. One student emphasized:
\begin{quote}
    ``\textit{College is not easy. You're taking high level classes trying to do life at the same time. There's just this constant pressure on you. I'm definitely more prone to AI as a college student, not just for time and benefit, but [because] struggling as a college student. I feel like I almost have to use AI to succeed, to graduate at the end of the day, to get to end of this class.}'' (P1)
\end{quote}
Students acknowledged their emotional states of stress and anxiety significantly influence the likelihood of using AI, as one student explained, ``\textit{Everyone is tired and anxious all the time. If I was less stressed, I would devote more time...before turning to AI}'' (P14). One other participant echoed, ``\textit{If I'm getting too stressed out and I can't figure it out, that's my turning point}" (P22). Mood also shapes the quality of students' engagement with AI. One student reflected, ``\textit{when I feel good, I put a lot more care...in the way I prompt AI. [But] on a day I was really depressed...I would just be very vague}" (P1). These accounts show persistent stress from college leads students more likely to delegate to AI for efficiency rather than intentionally use it. 

\subsubsection{Deadlines and Exam Cycles Intensify AI Use}
\label{deadlines}
Students reflected that their AI usage increased toward the end of academic quarters when approaching deadlines. Participants acknowledged, ``\textit{later in the quarter...that was when AI became more attractive}'' (P21). Factors such as \textit{``time crunch, stress, and pressure"} (P4) became immediate drives for AI use. Under pressure of multiple deadlines, students shifted from genuine trying to learn to task completion, as captured in this example: 
\begin{quote} 
``\textit{This is going to be due tonight. My deadline is in 10 minutes. I am still not grasping it. And I upload a file into ChatGPT and say, hey, can you create a code for the homework?} '' (P19) 
\end{quote}  
Across participants, they repeatedly described midterms, finals, and stacked deadlines as moments when they had ``no choice'' (P1) but to shift toward expedient AI use.

Interestingly, we noticed AI's accessibility sometimes encourage procrastination and shift time management when students faced with multiple deadlines. For example, knowing AI could serve as a last-minute rescue, one participant admitted: ``\textit{I feel like it makes me procrastinate more...I'll be sitting there and I'll be like, oh yeah, probably around 11 o'clock [the deadline is at 11:59 pm], I'll take a look at the reading and have AI just quickly summarize it down}" (P8). Another student described how AI has influenced their time management: ``\textit{But now I'm just like, let me just wake up at 12, let me chill out for a bit. And then at three, I'll just start it [an assignment]. I have AI, so it's all going to be easier}'' (P1).

\subsubsection{Grades Prioritize Output over Learning: AI for Validation and Pre-Grading Checks for Assignment}
Participants said wanting good grades as a major drive to use AI. One student said, ``\textit{we care more about the grade than what we're actually learning...this is one of the biggest factors in people using AI}'' (P19). The same participant elaborated that cultural emphasis on grades led students to ``\textit{trade my learning for a good grade}'' (P19). Their AI usage is propelled by an education system that evaluates and rewards output instead of the learning process.

Driven by the desire to achieve high grades, students used AI as a assignment validation tool, leveraging it to check their assignments against grading criteria before turning in, as one participant illustrated:
\begin{quote}
``\textit{I put the rubric and my assignment into AI to double check everything. It makes me confident submitting it, knowing I've checked everything off.}'' (P1)
\end{quote}
Another echoed, ``\textit{ask ChatGPT to make sure I've gotten every single bit of what the rubric is asking me}'' (P4). Students also described STEM homework such as coding tasks were often checked by AI for
correctness: ``\textit{For STEM classes, I ask ChatGPT if I'm correct}'' (P5), and ``\textit{I write the code myself and then I ask AI to check it}'' (P3). Across these cases, AI served as a ``\textit{reassurance tool}" (P4), ensuring students submitted assignment that were free of errors and could earn them good grades.

\subsubsection{Triaging Across Courses and Tasks: Offload Perceived Easy, Menial, and Low-Value Tasks to AI}
Faced with competing demands from various courses, students reported using AI for task triage purpose: they used these tools to manage their workload, often by offloading effort from lower-priority tasks to protect time for the perceived critical ones. For example, they said they were more likely to automate small, low‑point assignments: ``\textit{weekly assignments...I did try to walk through them, [but] they were very little points, so I just used ChatGPT to help me answer those questions}'' (P22). For electives or less valued courses, they also delegated to AI to finish homework: ``\textit{If the elective class, I really don't care...deadline an hour, I would just give it to the AI}'' (P17) and ``\textit{this one class...I'm just going to put this through ChatGPT}'' (P19). 

The tasks triage led students reported offloading to AI for tasks that were  perceived as low-value, easy, menial, yet time consuming. One student said, ``\textit{this feels a little bit like busy work, so I'm using AI}'' (P10). Others described delegating reading reflection assignment, saying, ``\textit{I already read the chapter...answer a couple of questions, maybe paraphrase...I'll maybe go into AI}'' (P1). Students framed this as rational triage: ``\textit{simple tasks like readings, I would just have AI handle it}'' (P5) and ``\textit{calculating things or boring automated tasks...that's fine}'' (P8). Students consciously evaluated assignments against their perceived value, compared urgency of multiple courses' requirement, and offloaded less critical tasks to AI to manage their overall workload.

\subsubsection{Student Suggestion: More In-person Quizzes and Exams}
Students imagine fundamental changes to assessment structures that acknowledge AI's presence. Several participants advocated for frequent low-stakes quizzes and evaluation with in-person exams. One student explained, ``\textit{By doing weekly low-stake quizzes, it's not only [put] less pressure and stress on the student, but it also encourages students to actually try to use AI [to prepare]...I actually tend to engage with the material better}'' (P9).

Others echoed this recognition of new forms of evaluation by suggesting ``have AI open for homework...but make the exams in-person" (P17). Another noted some courses already ``switched every exam to in person" (P6). A pre-doctorate instructor even proposed oral exams instead of assigning essays that could be supported by AI easily, saying ``\textit{students would hate that. But I mean...eloquently speak about something is an aspect of learning too. So I've been thinking about [switching to oral exams], but I'm afraid of the students showing up with pitchforks at my office.}" (P18).

\subsubsection{Student Suggestion: Reduced Pressure Enables Reflective AI Use for Learning}
Students also imagined more purposeful AI use when asked if they were not pressured by the aforementioned institutional factors. One said, ``\textit{I would use it to almost have fun and gain knowledge rather than do this or do that for me…the negative impact comes from that time constraint}'' (P1). Another hoped to ``\textit{prioritize what I'm getting out of the assignments rather than…score a certain percentage}'' (P2). Across participants, they mentioned pressure pushed them toward expediency, while less pressure might lead to more exploratory and learning‑oriented use.

\subsubsection{Student Suggestion: Argue for AI Literacy and Resources On Campus}
The majority of students were unaware of any campus resources about AI for learning, saying something like ``\textit{I haven't seen [resources about] how to use AI, how to be mindful of it. No, I haven't seen anything like that}'' (P8), and ``\textit{I'm sure they're there, but I couldn't name one}'' (P18). One student assumed there were some resources but could not accurately name where to find them, ``\textit{I'm sure they're there, but I couldn't name one…if someone asked me for resources on AI, I would probably point them to the library. But other than that, no.}''
Students consistently advocate for AI literacy classes on campus. They asked for practical instruction in prompt engineering, ethical use, and critical evaluation of AI outputs. One student framed this as a matter of equity: ``\textit{People deserve the right to know what AI is}'' (P6). The current absence of school-wide support leaves students to navigate AI use independently without guidance.

To summarize, these environmental and institutional factors influence students for AI use by rewarding speed and output under academic performance pressure. Students' own suggestions show how relief from those pressures could lead to meaningfulengagement with AI toward learning goals. They also argue for more in-person quizzes and exams for better assessment structure.

\subsection{The Influence of Social Pressure and AI Shame}

Students described their immediate peer communities as the primary context that determined their adoption and usage of AI, often overriding AI policy and teachers' instructions. However, despite widespread adoption on campus, explicit discussion of AI remained taboo and created the phenomenon called ``\textit{AI Shame}", building an environment of implicit rules and hidden practices.

\subsubsection{Immediate Communities Set De Facto Local AI Usage Norms}
Micro-communities such as friend groups and sports teams shaped the acceptance of AI usage through shared accountability and norms. One student described how their hockey team's shared accountability and team morale reduced AI use:
\begin{quote}
``\textit{A lot of us understand if we get caught using AI irresponsibly, that not only affects school, but that affects our hockey team. So a lot of us were very careful using it or just didn't use it at all.}'' (P7)
\end{quote}
For this student, social liabilities within a close relationship group directly reduced AI adoption. 

Conversely, students often mentioned adopting AI because everyone else in their immediate peer group was using it, even if it was against the AI policy set up by their instructors. Peer influence, described by participants as ``\textit{a hundred percent}" responsible for their AI adoption, commonly spread through word-of-mouth (P21). They say something like ``\textit{because I saw my entire community using it},'' (P2) ``\textit{because other people were using it and describing how they found AI helpful},'' (P16) and ``\textit{I've had professors who completely disapproved of using AI...But then if I see my friends being able to get accurate answers and be able to understand things better with AI, then I'm more encouraged to use it}'' (P8).

Students expressed fear-of-missing-out (FOMO) feelings, worrying they would fall behind peers if they did not use AI. As one student noted during a group study session ``\textit{one person is using ChatGPT and racing ahead...my competitive nature crept in...[If I don't use AI,] I'll feel FOMO}'' (P4). And another one stated ``\textit{everyone else in my program was using it}'' and not using it meant ``\textit{making your life harder}'' (P21). Others described a perceived disadvantage for not using AI: ``\textit{you have to assume everybody uses it even though the syllabus doesn't allow},'' so not using it puts you ``\textit{at a disadvantage}'' because peers ``\textit{fact check their work or understand concept better than if you were to not use AI at all}'' (P2). Even with disapproving instructors, seeing friends ``\textit{get accurate answers}'' encouraged use (P8). 

This shows students' engagement with AI relies significantly on their immediate community norms and the feeling of FOMO rather than school policies or teachers' rules.

\subsubsection{An Open Secret: The Pervasive Culture of AI Shame}
Despite the widespread adoption of AI tools across students on campus, participants described discussion of AI use is still a taboo topic. As one participant said with frustration, ``\textit{everyone's using it, but no one is allowed to talk about it}'' (P21). Another called it ``\textit{a hidden topic...a forbidden topic}'' that most would not discuss openly (P6), while another mentioned an interesting phenomenon on campus:
\begin{quote}
    ``\textit{I've seen so many people that sit in front of me in different classes. Whenever they switch to chat GPT, they lower their screen brightness to minimum…If you are dimming the screen so low, then sure, I'm assuming that you're copy and pasting that quiz into AI. It's so funny...There's that associated feeling of shame or guilt when using AI in class}'' (P6). 
\end{quote}
Such implicit culture created an atmosphere where nearly universal AI usage remained covert due to AI shame and potential reputational risk. Students worried that open AI use could signal laziness or dishonesty. One said, ``\textit{it feels like cheating}'' (P3); another felt ``\textit{extremely guilty}'' after using it daily (P18). Some framed the stigma ``\textit{there is a shame...maybe taboo}'' and ``\textit{a little bit of embarrassment},'' but they did not think it should be hidden (P9). Reputation concerns limited disclosure: ``\textit{if I admit to using it...I'll have a reputation as not as good of a worker...I only talked about how I used it with really close friends}'' (P21). Sometimes AI shame led to suspicion and wrongful accusation, as one student described a friend was wrongly accused of using AI for a group project because she had bullet points in the writing which was stylistically similar to AI output. Other project teammates ``\textit{wanted [the friend] to transform the whole document}'' to avoid an AI look, ironically because ``\textit{themselves [the teammates] are using AI but they don't want their assignment to look like it}'' (P13). These moments show AI shame is pervasive in college, pressuring students to be cautious of their work's resemblance to AI output, even when the work itself is not supported by AI. And the quotes show prevalent stigma fosters widespread hidden practice of using AI on campus. 

\subsubsection{Student Suggestion: Perceived Openness in Professional Fields and Call to Destigmatize AI Shame on Campus}
Participants mentioned they were aware of a clear divide between academic and professional environments regarding AI openness. Compared to the hidden atmosphere in academia, professional environment were perceived much more welcoming, as one participant said, ``\textit{Companies are definitely shoving it down everyone's throats since there's less tension socially}'' (P16). Students advocated normalizing meaningful use: ``\textit{I don't think it's that bad of a thing}'' and as people learn ``\textit{how to effectively use it, everyone started to embrace it}'' (P4). Others urged the campus to ``\textit{be more explicit}'' about AI culture and stop ``\textit{avoiding this topic},'' calling the silence a ``\textit{waste of opportunities}'' to learn both potentials and risks (P6). Thus, participants criticized college institutional avoidance of AI discussions, advocating for explicit acknowledgment and normalization to mitigate the secrecy.

\subsection{The Influence of AI Policies}

Students described a landscape of inconsistent AI policy patchwork on campus. These unclear rules made them feel confused, anxious, and often they went against the policy in practice. 

\subsubsection{Inconsistent and Outdated Policies Breed Confusion and Anxiety}
Participants often reported institutional AI policies as unclear, inconsistent, and outdated. It was difficult for students to build consistent understanding of these policies, as one student bluntly said, ``\textit{It's a mess. Different classes have different requirements. It's really up to the instructor and how they design their courses}'' (P6). Students faced varying standards with one class allowed AI for research, another banned it entirely, while yet another required attribution (P16). Even instructors felt lost as reflected by a PhD student who was also a pre-doctorate instructor, ``\textit{I am a teacher and I don't know what the policy is…I don't know how to enforce it}" (P18). These policies confusion caused frustration and interpretation ambiguity, making students anxious of unintended violations. One participant withdrew from a course after misunderstanding vague policy language, fearing accusations of plagiarism:
\begin{quote}
"\textit{My instructor clarified that students using AI had been forwarded to academic review. Although I didn't plagiarize, and I genuinely did check the syllabus to understand what was expected of me. [But] I realized that I misinterpreted the syllabus...and for the sake of my own academic integrity and also not causing any further difficulties with the school, I chose to withdraw from the class}" (P15).
\end{quote}
Students repeatedly called for clearer AI policies across courses to mitigate confusion: ``\textit{There is a bigger need for stronger tech policy for AI. But for now, it's heavily dependent on students' interpretation}" (P9). Currently, the inconsistent policies shifted responsibility onto students for their own interpretation, causing stress and fear of misunderstanding.

\subsubsection{Generic Policies Lead to Student Disengagement}
Many students also said courses' AI policies as generic, and perceived these policies as instructors copy and paste from other classes without much thought. They mentioned common wording like ``\textit{don't copy and paste answers}'' (P5), ``\textit{don't use it on a test}'' (P7), ``\textit{don't cheat...and attribute use}'' (P10), ``\textit{grammar check is fine, not idea generation}'' (P12) frequently appeared in syllabi. Students dismissed these policies as redundant and not specific, consequently leading them to overlook the policies. For examples, students said they ``\textit{don't pay too much attention}'' to AI sections (P9), ``\textit{don't think they [students] read the policy rules anyways}'' (P3), and ``\textit{zone out when they're talking about AI policies}'' (P4). Even instructors admitted copying AI policy wording from other courses and doubted its relevance, as the pre-doctorate instructor mentioned above said, ``\textit{I just copied it from another syllabus}'' (P18). The instructor also mentioned a discussion with students instead of just listing generic policies on syllabus would align teacher-student objectives, ``having a conversation about it…I want to convey to them the way I grade is based off not about strength of writing. It's if I can hear your personal experience" (P18).

\subsubsection{The Normalization of Policy Violation}
Participants described widespread violation of AI policy is common on campus. When asked if classmates go against policy, one answered, ``\textit{Definitely...lots of people did}'' (P21). Another emphasized the scale of violation, ``\textit{many friends got in trouble},'' and the professor ``\textit{wrote up half of the class for AI use}'' (P2). One participant mentioned used an AI detector to check peers' discussion post answers on Canvas \cite{canvas_2021} (a learning management system that students can post answers to discussion questions online), finding ``\textit{80 to 90 percent}'' flagged as AI-generated (P13). Students choose to violate policies when under pressure, saying ``\textit{if the homework is due in an hour...students still use AI}'' even when it is not allowed (P17). Because detection tools exist, a participant said students need to ``\textit{tread with AI cautiously}'' to managing risk rather than avoiding violation (P3). The participant then summarized the reality on campus starkly: ``\textit{the policies...don't [work] to the majority}'' (P3). This widespread noncompliance suggests that current policies fail to align student practices with institution's policy goals, creating an environment where violation becomes the norm rather than the exception, and students develop tactical strategies to circumvent AI detection.

\subsubsection{Student Suggestion: Rejecting Blanket Bans and Suggesting Discussion and Co-Created AI Policy}
When asked about their ideas for developing better AI Policy on campus, students said they did not think blanket bans of AI usage would work, and argued that simple bans might increase students' anxiety and push the use underground. One participant said an AI ban, ``\textit{is harmful in that it creates an atmosphere of anxiety around using it even in positive circumstances}'' (P15). Another noted, ``\textit{to make it stricter would further encourage students to use AI}'' (P9).

Instead, students advocated for co-creating AI policies with instructors that build understanding, awareness, and accountability. As one participant proposed:
\begin{quote}
``\textit{I would love it if the class collaboratively comes up with a technology use policy for us to all follow...What does this look like? How often should we use it, when should we not use it and why? I definitely want it to be something that we come together as a class to decide on}'' (P23).
\end{quote}
Students also emphasized setting clear AI usage expectation at the beginning of the class, ``\textit{It would be good to have a brief discussion at the beginning of the course of examples of what's acceptable and what's not acceptable}'' (P16), and another suggested an a lightweight quiz to make sure students were clear of the AI rules (P18). 

\subsubsection{Student Suggestion: Department-Level Policy Making with Course-Specific Flexibility}
Students recognized that AI policies cannot be one-size-fits-all across disciplines. They call for department-level policy that considers disciplinary differences while providing consistency within each department. One participant argued: 
\begin{quote}
``\textit{It's probably a good time for the department to gather all instructors to reevaluate their courses and make a one-page of an AI policy for the department...every department is going to be different. That careful discussion and consideration should been done}'' (P6).
\end{quote}
Another student echoed, ``\textit{I think it would be important for each different program to discuss that and decide it for themselves}'' (P21). Students also understand course differences should be considered: ``\textit{A class about learning how to write versus a class about learning how to code is really different}'' (P2), suggesting policies should reflected specific learning objectives of each course.

\subsection{The Influence of Internal Values and Learning Goals: Students' Strategies for and Struggles with Self-Regulating AI Usage}

Students showed a clear awareness of AI's potential to foster superficial learning. They tried to police themselves into engaging responsibly with AI by implementing value-based self-regulation strategies to prevent superficial learning. However, external pressures frequently weakened their ability to maintain these self-imposed rules, leading to the ``intention-behavior gap.''

\subsubsection{Awareness of Superficial Learning with AI}
Many participants admitted that using AI often led to superficial learning and poor retention. As one participant said, AI use often felt ``\textit{more superficial},'' and another reflected, ``\textit{When I use AI, I haven't seemed to remember anything. I don't think I retained it well enough or learned deeply enough}'' (P22). One participant described AI undermining the learning process, saying, ``\textit{AI picks me up from point A and drops me at point B without me learning the journey. Then when I try to get to point C alone, I don't know how. I'm stuck without the learning experience}'' (P19). This understanding led students to consider how to regulate their usage of AI to preserve learning opportunities.

\subsubsection{Self-Regulation Rooted in Students' Core Values}
To prevent these superficial learning experiences, students developed self-regulation strategies based on their values. These values included valuing critical thinking, creativity, and the cognitive struggle required for deep learning. The specific values and self-regulation strategies students mentioned: 

\begin{itemize}
    \item \textit{\textbf{Value the process of thinking}}: ``\textit{I really enjoy the process of me thinking through my daily learning experience and figure out stuff...I don't think I want it [AI] to replace my thinking process.}" (P6). The participant added, ``\textit{I don't want to see ideas [from AI] first because it will skew my thinking}" (P10).
\end{itemize}
\begin{itemize}
    \item \textit{\textbf{Value creative ideation process}}: ``\textit{There is value to coming up with ideas on your own}" (P16). Another student echoed ``\textit{If I have to think of something creative, I would try to think first}'' (P5). One other student limited their use ``\textit{for creativity purposes,}'' and admitted, ``\textit{It makes me feel that self-discipline is very important in this space}" (P9).
    \item \textit{\textbf{Value learning struggle and willpower}}: ``\textit{It is about the willpower to say no I will sit and I will struggle}" (P7). Others described learning was meant to be uncomfortable, ``\textit{in the way that you are meant to sit down and think about things and process difficult concepts in your head and go through that difficulty, sit with that discomfort}" (P10). One student purposefully avoids AI on practice exams and ``\textit{struggle...almost a whole day}" before returning to it (P22).
    \item \textit{\textbf{Value critical thinking}}: ``\textit{I've had to think really critically about how to use it}" (P21). Another insisted, ``\textit{I truly think that [for] the process of critical thinking and ideation, we should be coming up with our own ideas,}" and that AI helps only ``\textit{if you are actually analyzing what it gives you}" (P18).
    \item \textit{\textbf{Value personal integrity and growth}}: ``\textit{When I decide not to use it for writing...it is within my values: personal integrity, academic integrity, and personal growth. I think it [using AI] is a shortcut}" (P21). Another described an ``\textit{invisible framework}" based on their value system for integrity use (P15).
\end{itemize}

These values serve as a foundation for students to make decision about if and how they engage with AI.

\subsubsection{Developing Metacognitive Skills for AI Use}
Students also recognized that learning to use AI effectively is itself a metacognitive skill requiring learning. It includes understanding when and how to effectively prompt AI, evaluating its outputs, and maintaining alignment with their educational objectives. One participant described this learning process: ``\textit{It actually is challenging to use in some ways because you have to know what your goals are and how you learn and how you think}'' (P21). 

The metacognitive work involves constant evaluation of when AI helps versus detracts learning. Students mentioned they were actively developing their learning metacognition, as one reported developing their rule of thumbs: ``\textit{If I know I can do it, then I'm more open to using AI for it. But if I don't understand it at all or don't know what to do, then I don't agree with using AI because I'm not learning anything}'' (P8). Another participant summarized the need for meta-awareness in different leaning contexts: ``\textit{I could accidentally use AI irresponsibly...it would be very dangerous if I don't know I'm in that situation [coding context]}" (P6). These reflections show learning with AI put more emphasis on developing students' metacognition during the learning process. 

\subsubsection{The Intention-Behavior Gap Under Pressure}
Despite reported intentions to self-regulate AI usage in learning context, students consistently experienced a gap between their ideals and actual behavior. One participant captured this internal conflict: ``\textit{I still feel really torn about it. I feel ethically, morally, and just personally uncomfortable using it, but it's so useful. It's hard not to}'' (P20). 

Students explicitly connect this intention-behavior gap \cite{sheeran2002intention} to competing priorities and time pressures (as detailed in section \ref{deadlines}): ``\textit{My goals right now are different than the habits that I want to develop in the future...While juggling multiple classes, multiple clubs, multiple outside projects, I have a lot of stress. And so I choose whatever's easy and what can get things done properly and quickly}'' (P9). This behavior-intention gap shows the vulnerability of students' self-regulation, and individual willpower alone is not sufficient to counteract the temptation of using AI under pressures.

Across participants, students clearly recognize the costs associated with superficial AI use and actively attempt to self-regulate rooted in their values. However, institutional pressures regularly detract these individual intention and efforts.

\section{Discussion}
Based on our results, we discuss what higher education institutions, instructors, and system designers can do to support student learning in a world where AI is readily accessible.
\subsection{Higher Education Institutions: Rethinking Assessment and Policy, Fostering AI literacy and Openness}
As generative AI becomes increasingly integrated into student workflows, higher education institutions face a critical need to evolve their practices. Our findings show a growing misalignment between students' AI adoption and existing institutional practices, notably around assessment, policy, and limited AI resource. This misalignment has led to covert AI usage, increased student anxiety, and an intention-behavior gap where students feel pressured into dis-regulated AI engagement. Institutions should shift from developing policies focused on plagiarism \cite{simkute2025new, pudasaini2024survey, mogavi2024chatgpt, huallpa2023exploring, rajabi2023exploring} and instead rethink how assessments are conducted, implement better policy architecture, build AI literacy and access, and destigmatize AI use.

\subsubsection{Rethinking Assessment Practices for an AI-Integrated Context}

Our results show that traditional assessment structures---characterized by tight deadlines, exam stacking, and graded homework---incentivize students to use AI for task completion rather than deep engagement. Our findings indicate that students frequently use AI to validate their homework against rubrics, prioritizing good grades. This behavior aligns with prior research, which shows that formulaic assignments and short-form essays are highly susceptible to superficial AI use \cite{rudolph2023chatgpt}. Building on literature \cite{simkute2025new, smolansky2023educator, rudolph2023chatgpt}, we suggest assessment can shift towards including reflection on learning objectives, explicit justifications of AI usage, encouraging collaboration, and emphasizing critical thinking to counter the effect posed by individual interaction with AI. In addition, as students themselves suggested, low-stakes quizzes, in-class exams, and oral exams could reinforce knowledge acquisition and prompt students to engage with the learning material.  

Moreover, integrating AI literacy into evaluation criteria \cite{rudolph2023chatgpt}, such as assessing critical prompting skills, verification AI output skills, and metacognitive reflections on AI use  \cite{knoth2024ai, lee2025prompt}, would empower them for future professional expectations, where AI integration is widely accepted \cite{portocarrero2025artificial}.

\subsubsection{Proposing A Three‑Tier AI policy Architecture}

Our research identified widespread student confusion stemming from inconsistent, vague, or generic AI policies across disciplines and courses. Students described significant anxiety and stress due to the burden of interpreting varied AI guidelines, resulting in unintended noncompliance or stealthy use. Blanket bans on AI were frequently viewed by students as unrealistic and counterproductive, aligning with prior studies suggesting banning is unsustainable and detrimental to cultivating AI literacy \cite{rudolph2023chatgpt, kelly2023chatgpt, chan2023comprehensive, luo2024critical}. 

Our findings suggest that there is a need for multiple levels of AI policy. For example: 1) At the campus level, institutions could establish clear definitions and standardized vocabulary related to AI use. For example, defining terms such as AI-assisting (AI contributes ideas, structure, or information), AI-editing (AI checks spell/grammar, rephrase sentence style), AI-guide (AI gives hints and tips but does not offer solution directly, similar to Google Gemini's Guided Learning features \cite{heymans_2025}). These definitions should delineate different levels of permitted engagement, alongside explicit disclosure standards, examples and templates, and scaffolding to help students understand privacy guidelines. 2) At the departmental level, programs can publish short discipline-specific policy consensus, developed collaboratively with students and faculty, to ensure consistency and relevance within disciplines. 3) At the course level, instructors could tailor policy details to specific course objectives, co-creating these policies with students through participatory design methods \cite{spinuzzi2005methodology}.

\subsubsection{Fostering AI Literacy and Destigmatizing AI }

Our findings indicate that students have limited awareness and access to campus AI literacy resources, often navigating tool usage based on their own experience rather than guidance. In addition, previous research has found that this disparity could be exacerbated by students who can afford paid AI subscriptions versus those that cannot \cite{francis2025generative}. Institutions need to establish structured AI literacy programs and courses that provide enough resources to all students, as suggested in recent literature \cite{johnston2024student, kong2021evaluation, chan2023students}. Schools can consider having equal baseline access to tools through campus-wide licensing or credits. 

Additionally, our findings illustrate a culture which students feel compelled to hide or modify AI-generated work. Such covert practices create an implicit, peer-influenced drive of using AI regularly but secretly, undermining transparency and equity in the classroom. To encourage transparency, school advisors can host workshops, discussions, and engage students to explore constructive ways of using AI. In addition, to break the current micro-community-based AI norms and respond to the rapid evolution nature of AI technologies, institutions can also facilitate student-led knowledge-sharing. Encouraging students to share evolving best practices may maintain relevant resources more sustainably than institutionally managed content.

\subsection{Instructors: Aligning Assignments with Explicit Learning Goals, Facilitating Group Norms}
Currently, learning objectives are often implicit and students usually are not aware or pay attention to. However, if students understand these objectives well, they can develop metacognition and self-regulation of AI use to align with these goals, rather than passively relying on institutional policies. Thus, communicating these objectives to students is critical. For example, teachers could use Bloom's learning objectives \cite{adams2015bloom, chatterjee2017write} to educate students when AI use is beneficial or detrimental for specific learning outcomes. For instance, if the learning goal is recall and remembering, instructors may implement more retrieval‑practice such as low‑stakes quizzing \cite{kenney2021low} to build memory without AI shortcuts. Or instructors can communicate to students the learning objectives mentioning they could use AI as learning guide (such as building flashcards). For the recalling learning objective, teachers should clearly dissuade them from using AI irresponsibly, explaining that dependence on AI could negatively impact long-term retention and conceptual understanding \cite{bastani2024harm}. 

Additionally, to address the social influence of AI usage, instructors can facilitate group projects by forming teams rather than relying on self‑selection and by setting explicit team‑level AI norms. Random or instructor‑formed teams can prevent micro‑communities from creating inequitable norms \cite{oakley2004turning}, and is critical since our study shows students' immediate groups set the AI usage norms regardless of policy. Also, instructors can encourage and facilitate groups to set team charters which could include how AI is permitted in the group project. Literature shows clear, shared norms can also leverage peer influence toward integrity rather than hidden practice \cite{mccabe1997individual}.

\subsection{System Designers: Incorporating Learning Goal Affordance, Supporting Instructors to Make Policy}

AI tool designers can facilitate learning goal oriented use by explicitly incorporating both student's and instructor's learning and teaching goals into system functionalities. Systems could let instructors set specific parameters around students' AI assistance, exposing controls to configure what kinds of assistance are appropriate at different times and for different tasks \cite{holstein2022designing}.

In addition, consistent with self‑regulated learning \cite{winne2021open}, systems should allow students to input their own learning objectives, allowing AI systems to adjust support based on them. For example, AI system can support self-regulation by supporting different AI usage decision points (e.g. editing-only mode or an idea-generation mode depending on the user's specified learning objective).

Some universities have already started to implement webpage and systems to support instructors to set up their AI policies \cite{stanford_2025, ut_austin_2024}. To build upon these initiatives, AI policy makers and system designers could collaborate to develop systems that support instructors in generating customized AI policies aligned with specific curricula and teaching preferences. An ideal solution would enable instructors to input course-specific details such as learning objectives, course content, and desired learning outcomes, then receive tailored AI policy recommendations and practical implementation strategies. Such a system would reduce the current burden on instructors.

\section{Limitations and Future Work}
Our study has several limitations, including a relatively small sample drawn from one large public university in the United States, limiting generalizability across different institutional contexts, cultures, and regions. Future research should include larger and more diverse samples to capture a broader range of environmental influences. Additionally, since all participants were already users of AI tools, perspectives from non-users are absent. Understanding non-users’ views and experiences on campus would further reveal institutional and social impacts on student AI usage. Lastly, our study took place between May and July 2025, prior to some technological updates such as Gemini's Learning Guide and ChatGPT's Study Mode. As technological capabilities rapidly evolve, future research should continually explore how students respond to new technology developments and shifting campus cultures.

\section{Conclusion}
Our study reports how institutional pressures, social influences, and inconsistent policies shape students' generative AI usage in higher education. Students frequently rely on AI under deadline and grading pressures, influenced strongly by peer norms despite formal AI guidelines. Students actively develop self-regulation strategies to align AI use with their values but often face an intention-behavior gap under persistent external pressures. Students also suggest collaborative, course-specific AI policies, increased in-person evaluations, and AI literacy programs. Our findings emphasize the need for institutions, instructors, and tool designers to collaboratively create environments that support integration of AI into student learning experiences.

\bibliographystyle{ACM-Reference-Format}
\bibliography{references}
\appendix*
\section{Appendix A: Participant Table}
\label{Appendix:1}

\begin{table}[t]
\centering
\caption{Participant demographics and reported GenAI tools used (N=23). Interviews were conducted with undergraduate, master’s, and doctoral students at a large public university in the United States. $^\ast$Major column shows anonymized participants' specific programs (anonymized to STEM and Non-STEM) for manuscript submission and review.}
\label{tab:participant_demographics}
\renewcommand{\arraystretch}{1.2}
\rowcolors{2}{gray!10}{white}
\resizebox{\textwidth}{!}{%
\begin{tabular}{ l >{\raggedright\arraybackslash}p{4.4cm} >{\raggedright\arraybackslash}p{2.7cm} l >{\raggedright\arraybackslash}p{8.0cm} }
\toprule
\textbf{Participant} & \textbf{Major (Anonymized)$^\ast$} & \textbf{Degree Program} & \textbf{Year of Study} & \textbf{GenAI Tools Used} \\
\midrule
P1 & STEM & Bachelor’s & Junior & ChatGPT, Gemini, Grok, Claude, Copilot, Grammarly GO, Quillbot, Canva \\
P2 & STEM & Bachelor’s & Senior & ChatGPT \\
P3 & STEM & Bachelor’s & Senior & ChatGPT \\
P4 & STEM & Bachelor’s & Sophomore & ChatGPT, Gemini, Perplexity AI, Quillbot \\
P5 & STEM & Bachelor’s & Sophomore & ChatGPT \\
P6 & STEM & Bachelor’s & Junior & ChatGPT, Gemini, Copilot, NotebookLM, Deepseek \\
P7 & STEM & Bachelor’s & Senior & ChatGPT \\
P8 & STEM & Bachelor’s & Sophomore & ChatGPT, Copilot \\
P9 & STEM & Bachelor’s & Junior & ChatGPT, Gemini, Claude, Perplexity AI, Cursor, Character.ai, DeepSeek \\
P10 & STEM & PhD & 5th Year & ChatGPT, Gemini, Claude, Copilot \\
P11 & STEM & Bachelor’s & Junior & ChatGPT, Gemini, Perplexity AI, Grammarly GO, Canva \\
P12 & Non-STEM & Master’s& N/A & ChatGPT, Midjourney, Notion AI, Canva \\
P13 & Non-STEM & Master’s & 1st Year & ChatGPT, Midjourney, Stable Diffusion, Deepseek \\
P14 & STEM & Bachelor’s & Junior & ChatGPT, Gemini \\
P15 & Non-STEM & Bachelor’s & Senior & Canva \\
P16 & STEM & Bachelor’s & Senior & ChatGPT, Copilot, Canva \\
P17 & STEM & Bachelor’s & Senior & ChatGPT, Gemini, Claude, Perplexity AI, Copilot, Grok, Github Copilot \\
P18 & Non-STEM & PhD & 3rd Year  & ChatGPT \\
P19 & Non-STEM& Bachelor’s & Sophomore & ChatGPT \\
P20 & Non-STEM & Master’s & N/A & ChatGPT, Claude \\
P21 & Non-STEM & Master’s & 2nd Year & ChatGPT, Claude \\
P22 & Non-STEM & Bachelor’s & Freshman & ChatGPT,Copilot,Notion AI \\
P23 & Non-STEM & Master’s & 1st Year & ChatGPT \\
\bottomrule
\end{tabular}%
}
\end{table}

\end{document}